\documentclass[showpacs,twocolumn,preprintnumbers,amsmath,amssymb,aps]{revtex4}
\usepackage{graphicx}
\usepackage{dcolumn}
\usepackage{float}
\usepackage{caption}
\usepackage{bm}
\begin{document}
\title{Dissipative Quantum Systems and the Heat Capacity Enigma}
\author{S. Dattagupta, Jishad Kumar, S. Sinha and P. A. Sreeram}
\affiliation{ Indian Institute of Science Education \& Research-Kolkata,\\
Mohanpur, Nadia 741252, India}
\date{\today}
\begin{abstract}
We present a detailed study of the quantum dissipative dynamics of a charged
particle in a magnetic field. Our focus of attention is the effect of
dissipation on the low- and high-temperature behavior of the specific heat at
constant volume. After providing a brief overview of two distinct approaches to
the statistical mechanics of dissipative quantum systems, viz., the ensemble
approach of Gibbs and the quantum Brownian motion approach due to Einstein, we
present exact analyses of the specific heat. While the low-temperature
expressions for the specific heat, based on the two approaches, are in
conformity with power-law temperature-dependence, predicted by the third law of
thermodynamics, and the high-temperature expressions are in agreement with the
classical equipartition theorem, there are surprising differences between the
dependencies of the specific heat on different parameters in the theory, when
calculations are done from these two distinct methods. In particular, we find
puzzling influences of boundary-confinement and the bath-induced spectral cutoff
frequency. Further, when it comes to the issue of approach to equilibrium, based
on the Einstein method, the way the asymptotic limit ($t\rightarrow\infty$) is
taken, seems to assume significance.
\end{abstract} 
\maketitle
\section{Introduction}
Recent years have seen great strides in the statistical mechanics of dissipative
quantum systems \cite{1}. Dissipation arises when the quantum degrees of freedom
of a heat bath, which is strongly coupled to a subsystem of interest, are
projected (or integrated) out of the Hilbert space of the total system. Two
different approaches, detailed below in Sec.II, have been used in this context:
(i) the usual Gibbs approach that focuses on the partition function \cite{2} and
(ii) the Einstein approach that hinges on a quantum Langevin equation for the
subsytem \cite{3}. Lately it has been argued that the presence of quantum
dissipation yields a satisfactory behavior of the fundamental thermodynamic
attribute, viz., the heat capacity, as far as the low-temperature properties are
concerned \cite{4}. Here we will point out that there are some puzzling issues
even for the high temperature limit of the heat capacity, apart from the
intriguing low-temperature attributes. Before we address this question, it is
important to review the kind of subsytem we have in mind and the foundational
basis of statistical mechanics, which we do below. While our present
discussion as well as that in Sec.II are set within the domain of classical
statistical mechanics, extension to quantum mechanics can be easily carried out,
as indicated in Sec.III. But we want to first concentrate on some preliminaries
about the subject of statistical mechanics itself.

Statistical Mechanics provides the microscopic basis of the macroscopic
poperties of a system described by the subject of thermodynamics. Though the
power of statistical mechanics comes to the fore in its full glory for an
interacting many body system, such as in the exact formulation of second order
phase transitions by means of the two-dimensional Ising model \cite{5}, many of
the intricacies can be elucidated for just a single entity, albeit in contact
with a heat bath comprising an infinitely large number of (invisible) degrees of
freedom. It is this simplified approach to statistical mechanics in the context
of a single particle embedded in a heat bath that we shall adopt in this paper.

The dynamics of a particle of mass $m$ is described by the system Hamiltonian
defined by
\begin{equation}
 \mathcal{H}_{S}=\frac{\vec{p}^2}{2m}+V(\vec{q})~,
\end{equation}
where $\vec{p}$ is the canonical momentum vector of the particle moving under an
arbitrary potential $V(\vec{q})$ which is a function of the generalized
coordinate vector $\vec{q}$. We shall discuss three distinct cases in the
sequel: \vspace{0.5cm}

(a)\underline{Free particle}:
\begin{equation}
V(\vec{q})=0~,
\end{equation}

 (b)\underline{Harmonic oscillator}:
\begin{equation}
V(\vec{q})=\frac{1}{2}m\omega_0 ^2 \vec{q}~^2~,
\end{equation}
$\omega_0$ being the frequency of the oscillator, and \vspace{0.5cm}

(c)\underline{Charged oscillator in a magnetic field}, that is described by a
momentum and coordinate-dependent potential:
\begin{eqnarray}
V(\vec{q},\vec{p})&=&-\frac{e}{2mc}(\vec{p}.\vec{A}(\vec{q})+\vec{A}(\vec{q}
).\vec{p})\nonumber\\
&+&\frac{e^2}{2mc^2}\vec{A}~^2 (\vec{q})+\frac{1}{2}m\omega_0 ^2 \vec{q}~^{2}~,
\end{eqnarray}
$\vec{A}(\vec{q})$ being the vector potential, the curl of which yields the
magnetic field $\vec{B}$:
\begin{equation}
 \vec{B}=\vec{\nabla} \times \vec{A}(\vec{q})~.
\end{equation}
It is evident that for zero vector potential, case (c) reduces to (b). If
additionally, $\omega_0$ is also zero, case (a) is obtained. In what way are these
limiting situations arrived at, for a quantum dissipative system,  will indeed
be the focus of our discussion below. 

It should be mentioned here that the problem of a charged oscillator in a
magnetic field is relevant in the context of Landau diamagnetism \cite{6} which
has had a deep impact on modern condensed matter physics through phenomena such
as the quantum Hall effect \cite{7}. Landau diamagnetism, which is purely
quantum in origin, is characterized by strong boundary effects that can be
mimicked by the oscillator potential \cite{8}. The presence of a quantum bath,
comprising of, say, bosonic excitations like phonons, lends additional richness to
the problem as it allows us to study the effect of dissipation on Landau
diamagnetism \cite{9}. In this article however our focus of attention is not
diamagnetism but the thermodynamic property of the heat capacity.

The microstate of the particle at a given time is specified by a point in the
6-dimensional (three for coordinates and three for momenta) phase space. As the
time evolves the phase point curves out a phase trajectory. While in classical
mechanics the trajectory is uniquely deterministic,  once the initial values of
$\vec{q}$ and $\vec{p}$ are given, the point of statistical mechanics is that
the phase trajectory randomly changes from one `realization' of the system to
another. The meaning of `realization' becomes clear if one considers how
experiments are performed. A realization corresponds to a given experiment when
one watches the trajectory evolve in time. Of, course, the whole statistical
basis of  data collection is to repeat the experiment, this time tracking  a
different trajectory, even though the initial values of \{$\vec{q},\vec{p}$\}
are the same. It is this multitude of trajectories corresponding to multiple
realizations of the system that yields the concept of `ensemble' in statistical
mechanics $-$ an ensemble means a collection of possible realizations of the
system. Thermal equillibrium is said to be reached when experiments are repeated
so many times that all possible trajectories (realizations) in the phase space
are explored--this yields the notion of `mixing' \cite{10}.

With these preliminaries the outline of the paper is as follows. In Sec.II, we
review the Gibbs and Einstein approaches to statistical mechanics. Although our
treatments are couched in classical terms similiar results hold for quantum
phenomena as well. With these approaches in the background we summarize in
Sec.III, the newly developed subject of dissipative quantum systems. In Sec.IV
we analyze the results for the heat capacity for the three problems (a-c) and
point out certain surprises when we consider the various limits of case (c). In Sec.V, we summarize the results.
\section{Gibbs and Einstein Approaches to Statistical Mechanics}
The remarkable thesis of Gibbs is that for a system in thermal equilibrium the
observed properties of the system can be computed from a weighted average of the
values of the relevant observable at all possible phase points that lie on a
constant time-slice. This approach is quite different from how experimental data
are processed--by taking a time average of the `values' of the observable at
different times, over a very long time. The equivalence of this time-average to
the Gibbsian ensemble average follows from the fascinating attribute called
`ergodicity', a  property that is the consequence of mixing \cite{10}. The
ensemble average of an observable $X(\vec{q},\vec{p})$ in equilibrium (indicated
by the subscript `eq' below) is defined by
\begin{equation}
\langle X(\vec{q}, \vec{p}) \rangle _{eq} =Tr
\left(\rho(\vec{q},\vec{p})X(\vec{q},\vec{p}) \right)~,
\end{equation}
where `$Tr$'(trace) implies an integration over the entire phase space in
classical statistical mechanics, whereas it is a sum over possible eigenstates
of the full $\mathcal{H}_S$ in Eq.(1) in quantum statistical mechanics. The
Gibbs-Boltzmann weight function $\rho(\vec{q},\vec{p})$ is what is called a
density matrix, given by
\begin{equation}
\rho(\vec{q},\vec{p})=\frac{\exp(-\beta \mathcal{H}_S
(\vec{q},\vec{p}))}{\mathcal{Z}_S}~,
\end{equation}
where $\beta(=(k_B T)^{-1})$ is the inverse temperature, $k_B$ being the
Boltzmann constant. The normalization factor $\mathcal{Z}_S ~,$  referred to as
the partition function:
\begin{equation}
 \mathcal{Z}_S =Tr(\exp (-\beta \mathcal{H}_S (\vec{q},\vec{p}))~,
\end{equation}
provides the critical link between statistical mechanics and thermodynamics as
it leads to the Helmholtz free energy $\mathcal{F}$ through the relation:
\begin{equation}
 \mathcal{F}_S =-\frac{1}{\beta}\ln \mathcal{Z}_S ~.
\end{equation}
From $\mathcal{F}_S$ all thermodynamic properties can be derived.

It is of course outside the realm of Gibbsian statistical mechanics to address
the issue of how equilibrium is reached. That question has to be posed in terms
of models of nonequilibrium statistical mechanics, which are however not as
robust and time-tested as the formulation of equilibrium statistical mechanics
encapsulated by Eqs.(7)-(9). One model that stands out in this regard is based
on the idea of Brownian motion \cite{11}. In the latter one imagines the
particle (much like the pollen particle of Brown \cite{12}), the Hamiltonian of
which is given by Eq.(1), is in contact with a heat bath that drives stochastic
(noisy) fluctuations into the system. The idea of Brownian motion is very
physical in that  if one tags the particle by taking camera snapshots at
different times, its dynamics would indeed appear to be random, when the
particle is out of equilibrium, and even when it is in equilibrium! The
stochastic dynamics is captured by the time-dependant distribution function
$\mathcal{P}(\vec{q},\vec{p},t)$ in phase space that obeys the
Fokker-Planck-Smoluchowski-Kramers equation \cite{13}
\begin{eqnarray}
 \frac{\partial}{\partial
t}\mathcal{P}(\vec{q},\vec{p},t)&=&\{-\frac{\vec{p}}{m}.\vec{\nabla}_q+\vec{
\nabla}_p.(\vec{\nabla}_q V(\vec{q})+\gamma\vec{p})\nonumber\\
&+&m\gamma k_B T \nabla_p ^2\}\mathcal{P}(\vec{q},\vec{p},t)~,
\end{eqnarray}
where $\gamma$ is the friction constant. The quantity $\mathcal{P}$ plays the same role in non-equilibrium
as $\rho$ does in equilibrium. Thus the averaged time-evolution of the dynamical
variable $X(\vec{q},\vec{p})$ is given by
\begin{equation}
 \bar{X}(t)=\int d\vec{q}d\vec{p}
X(\vec{q},\vec{p})\mathcal{P}(\vec{q},\vec{p},t)~.
\end{equation}
With the temperature-dependant prefactor in front of $\nabla^2$, it is ensured that the stationary state is indeed the thermal equilibrium state, described by $\rho$ in Eq.(7). This is consistent with the fluctuation-dissipation theorem.

Although the fluctuation-dissipation relation is a necessary
condition for guaranteeing that the system transits to the thermal equilibrium
distribution, as $t\rightarrow \infty$, the Brownian motion model is far from
being a unique description for the approach to equilibrium. More significantly,
even within the Brownian motion model, there may be different routes to approach
equilibrium. For instance, we can ask: does
$\lim_{t\rightarrow\infty}\bar{X}(t)$ agree with $\langle X \rangle$, as defined
in Eq.(6)? The resolution to this question helps our understanding of how to
relate experimentally measured quantities to their theoretically calculated
values in equilibrium, as prescribed by Eq.(6), for instance (cf., comments in
the last but paragraph one in Sec.I).  Not surprisingly then, the rich physical
structure of the Brownian motion model has bestowed the latter the inspired
title of the `Einstein Approach to Statistical Mechanics'\cite{14}.

It is pertinent to mention here that the time-dependent approach,  as formulated
through Eq.(10), is based on what is called the `Schr${\rm \ddot{o}}$dinger
picture'.  An equivalent description obtains through the `Heisenberg picture' in
which one directly considers the dynamical equations of motion:
\begin{eqnarray}
 \frac{\partial \vec{q}}{\partial t}&=&\frac{\vec{p}}{m}~,\nonumber\\
\frac{\partial \vec{p}}{\partial t}&=&-m\omega_0 ^2
\vec{q}-\frac{e}{c}(\vec{q}\times\vec{B})\nonumber\\
&-&\gamma \vec{p}(t)+\vec{f}(t)~.
\end{eqnarray}
The set of equations (12) is called the Langevin equation in which the force
$\vec{f}(t)$ is a stochastic noise, defined on an ensemble for which the
distribution function is given by $\mathcal{P}(\vec{q},\vec{p},t)$. A particular
realization of $\vec{f}(t)$ corresponds to a given trajectory, and ensemble
averages are obtained by imposing the following constraints on the spectral
properties of $\vec{f}(t)$:
\begin{eqnarray}
 \langle \vec{f}(t) \rangle &=& 0\nonumber\\
\langle f_\mu (t) f_\nu (t^{\prime}) \rangle& =&2m\gamma k_B T \delta
(t-t^{\prime})\delta_{\mu\nu},~ \mu,\nu~=x,y,z~.\nonumber\\
\end{eqnarray}
\section{Dissipative Quantum Systems}
In this section we move from the classical to the quantal domain and consider
the case in which the quantum subsystem is put into contact with a heat bath
that is also quantum mechanical. Before we indicate the steps necessary for
Brownian motion in terms of what is referred to as the quantum Langevin
equations \cite{3},  it is useful to backtrack and indicate how the classical
Langevin equations (12) themselves are derived from a system-plus-bath method.
Here we start from a treatment of  Zwanzig \cite{15} in which the Hamiltonian in
Eq. (1) is extended as 
\begin{equation}
 \mathcal{H}=\mathcal{H}_S +\sum_{j}\left[\frac{\vec{p}_j ^2
}{2m_j}+\frac{1}{2}m_j \omega_j ^2 (\vec{q}_j -\frac{C_j \vec{q}}{m_j \omega_j
^2})^2 \right]~.
\end{equation}
Upon expanding the square over the round brackets it is evident that the
Hamiltonian contains a linear coupling between the coordinate $\vec{q}$ of the
subsystem and the coordinate $\vec{q}_j$ of the harmonic bath with $C_j$ being a
coupling constant.

From Eq. (14) it is easy to write down Hamilton's equations of motion, solve for
the bath coordinates and momenta, put the solutions back in the equations of
motion for the subsystem variables and derive for the momentum the generalized
Langevin equation \cite{13, 15}:
\begin{eqnarray}
 m\ddot{\vec{q}}&=&-m\omega_0 ^2
\vec{q}-\frac{e}{c}(\dot{\vec{q}}\times\vec{B})\nonumber\\
&-&m\int_{0}^{t}dt^{\prime} \dot{\vec{q}}(t^\prime)
\gamma(t-t^\prime)+\vec{f}(t)~, 
\end{eqnarray}
where the ``friction`` $\gamma(t)$, that appears as a memory function, depends
quadratically on $C_j$ and the noise $\vec{f}(t)$ depends explicitly on initial
coordinates and the momenta of the bath oscillators:
\begin{eqnarray}
 \gamma(t)&=&\sum_{j} \frac{C^2 _{j}}{m_j \omega_j ^2}\cos(\omega_j t)\\
\vec{f}(t)&=&\sum_j \{C_j [\vec{q}_j (0)-\frac{C_j \vec{q}(0)}{m_j \omega_j
^2}]\cos(\omega_j t)\nonumber\\
&+&\frac{C_j \vec{p}_j (0)}{m_j \omega_j}\sin(\omega_j t)\}~.
\end{eqnarray}
Suffice it to note that  Eq. (15) is exact and devoid of any assumption except
that we have decided to integrate the equation of motion in the forward
direction of time, thereby giving a sense to the `arrow of time'. The next step
however is a crucial one of introducing irreversibility by considering an
initial ensemble of states,  \textit{a}$^{\prime}$ \textit{la} Gibbs, in which
the bath variables are drawn at random from a canonical distribution (Eq.(7)),
yielding 
\begin{equation}
\langle f_\mu (t) f_\nu (t^{\prime}) \rangle = \delta_{\mu\nu}2mk_B T
\gamma(t-t^{\prime})~.
\end{equation}
The final step is to go to the limit of an infinitely large system in order to
endow the harmonic oscillator system the attribute of a heat bath. Thus
\begin{equation}
 \frac{1}{N}\sum_j C_j ^2 .... \rightarrow \int d\omega g(\omega),~m_j =m,~C_j
=\frac{C}{\sqrt{N}}~,
\end{equation}
where $g(\omega)$ is the `spectral density'. Equation (16) then yields
\begin{equation}
 \gamma(t)=\frac{C^2}{m}\int_0 ^\infty d\omega
\frac{g(\omega)}{\omega^2}\cos(\omega t)~.
\end{equation}

A commonly assumed form of $g(\omega)$  is the one which yields what is called
Ohmic dissipation, and is given by
\begin{eqnarray}
 g(\omega)&=&\frac{\omega^2}{\bar{\omega} ^3}~~~~~,\omega <\bar{\omega}
\nonumber\\
&=&0~~~~~~,\omega >\bar{\omega },
\end{eqnarray}
$\bar{\omega}$ being a high-frequency cut-off. Employing Eq.(21) we derive
Eq.(12), implying that Ohmic dissipation corresponds to \underline{constant}
friction $\gamma$ because the generalized friction coefficient reduces to
$\gamma \delta (t-t^{\prime})$, wherein $\gamma$ equals $\frac{3\pi
\hat{c}^2}{2m\bar{\omega}^3}$ \cite{13}.

The discussion in the quantum case proceeds along similiar lines in which one
has to however keep track of the fact that $\vec{q}$ and $\vec{p}$ are
non-commuting operators, and consequently, the noise $\vec{f}$ in Eq.(17) is
also a quantum operator \cite{3}. Additionally, because the bath oscillators are
to be treated quantum mechanically, the noise correlations are not 
\underline{`white'}, as in Eq.(13), but are characterized by both a symmetric
combination and a commutator structure, respectively given by
\begin{eqnarray}
 \langle \{f_\mu (t), f_\nu (t^{\prime})\} \rangle &=&
\delta_{\mu\nu}\frac{2}{\pi}\int_0 ^\infty d\omega \Re
[\vec{f}(\omega+i0^+)]\omega \coth(\frac{\beta\omega}{2})\nonumber\\
&\times& \cos[\omega(t-t^{\prime})]~.\\
\langle [f_\mu (t), f_\nu (t^{\prime})] \rangle &=&
\delta_{\mu\nu}\frac{2}{i\pi}\int_0 ^\infty d\omega \Re
[\vec{f}(\omega+i0^+)]\nonumber\\
&\times&\omega \sin[\omega(t-t^{\prime})]~.
\end{eqnarray}

At this point it is pertinent to ask: which system is $\beta$ (as in  Eq.(22))
the inverse temperature of ? In the Einstein approach, discussed so far in this
section, it is clear that $\beta$ represents the harmonic oscillator bath  which
the subsytem of interest, described by $\mathcal{H}_S$ in Eq.(1), is expected to
eventually come to equilibrium with. However, because the interaction between
the subsystem and the bath is treated exactly there is no reason for not
thinking of the entire system, represented by the Hamiltonian $\mathcal{H}$ in
Eq.(14), as one composite many body entity, which is further embedded in yet
another external bath, the inverse temperature of which is also given by
$\beta$! This then summarizes the Gibbsian approach in which one writes the full
partition function by replacing $\mathcal{H}_S$ in Eq.(8) by Eq.(14):
\begin{equation}
 \mathcal{Z}=Tr\left(exp[-\beta\mathcal{H}]\right)~.
\end{equation}
It is customary to rewrite $\mathcal{Z}$ as a functional integral \cite{16}:
\begin{equation}
 \mathcal{Z}=\oint \mathcal{D} [ \vec{q},\vec{p},\vec{q_{j}},\vec{p_{j}} ]
\exp\left(-\frac{1}{\hbar} \mathcal{A}_e [
\vec{q},\vec{p},\vec{q_{j}},\vec{p_{j} } ] \right)~,
\end{equation}
where $\hbar$ is the Planck constant and $\mathcal{A}_e$ is the so-called
Euclidean action, defined by
\begin{equation}
 \mathcal{A}_e =\int_0 ^{\hbar\beta} d\tau \mathcal{L}(\tau)~,
\end{equation}
$\mathcal{L}(\tau)$ being the Lagrangian written in terms of the `imaginary
time' $\tau (=i\hbar\beta')$. We illustrate in Sec.IV below the application of
Gibbs and Einstein approaches to the calculation of the heat capacity for the
charged oscillator in a magnetic field.
\section{Heat Capacity}
The heat capacity or the specific heat at constant volume is the most basic
thermodynamic property. It is defined by \cite{17}
\begin{equation}
 C=-k_B \beta^2 \left(\frac{\partial U}{\partial \beta}\right)_V ~,
\end{equation}
where $U$ is the internal energy. From a statistical mechanical point of view
$C$ is also related to the mean squared  energy fluctuations given by \cite{18}
\begin{equation}
 C=k_B \beta^2 \left(\langle \mathcal{H}^2 \rangle -\langle \mathcal{H} \rangle
^2 \right)~.
\end{equation}
While in the Gibbs approach $C$ can be directly computed from Eq.(25), employing
the definition in either Eq.(27) or Eq.(28), the quantities $\langle
\mathcal{H}^2 \rangle$ and $\langle \mathcal{H} \rangle ^2$ are functions of the
time $t$, in the Einstein approach. Correspondingly, $C$ will also be a function
of $t$, and the question we address is under what circumstances do we have the
following equality:
\begin{equation}
 \lim_{t\rightarrow \infty}{C(t)}^{{\rm Einstein}}=C^{{\rm Gibbs}}~?
\end{equation}
\subsection{Gibbs Approach ($\omega_0 \ne 0$)}
Before we discuss the calculation of  $C^{{\rm Gibbs}}$ for the dissipative
charged oscillator in a magnetic field it is useful to indicate the steps for
the simpler problem without dissipative coupling, viz; that described by
$\mathcal{H}_S$ alone (Eqs.(1) and (4)) \cite{19}. The corresponding Lagrangian
for the two-dimensional motion in the plane normal to the field is given by
\begin{equation}
\mathcal{L} = \frac{1}{2} m (\dot{x}^2 +\dot{y}^2 )-\frac{1}{2}m\omega_0 ^2 (x^2
+y^2)-\frac{e}{c} (\dot{x} A_x +\dot{y}A_y )~.
\end{equation}
It is customary to work in the so-called ''symmetric gauge`` in which 
\begin{equation}
A_x =-\frac{1}{2} yB,~~~ A_y =\frac{1}{2} xB~. 
\end{equation}
The Euclidean action can be written as
\begin{eqnarray}
\mathcal{A}_e [x,y] &=& \frac{m}{2} \int_0 ^{\hbar\beta} d \tau
[(\dot{x}(\tau)^2 +\dot{y}(\tau)^2 ) + \omega_0 ^2 (x(\tau)^2 +y(\tau)^2
)\nonumber\\
&-&i\omega_c (x(\tau)\dot{y}(\tau) - y(\tau)\dot{x}(\tau) )]~,
\end{eqnarray}
$\omega_c$ being the ''cyclotron frequency'' given by
\begin{equation}
 \omega_c =\frac{eB}{mc}~.
\end{equation}
Introducing 
\begin{equation}
x(\tau) = \sum_j \tilde{x} (\nu_j ) \exp( - i\nu_j \tau)~, 
\end{equation}
where $\nu_j$'s are the so called Matsubara frequencies,  defined  by 
\begin{equation}
\nu_j = \frac{2\pi j}{\hbar\beta} ~~~~ j=0,\pm 1,\pm 2,.... ~~~,
\end{equation}
we find
\begin{eqnarray}
\mathcal{A}_e [z_+ , z_- ]& =& \frac{1}{2} m\hbar\beta \sum_{j=-\infty} ^\infty
[(\nu_j ^2 +\omega_0 ^2 +i\omega_c \nu_j )\tilde{z}_+^*(\nu_j )
\tilde{z}_+(\nu_j )\nonumber\\
& +& (\nu_j ^2 +\omega_0 ^2 -i\omega_c \nu_j )\tilde{z}_-^*(\nu_j ) \tilde{z} _-
(\nu_j ) ]~,
\end{eqnarray}
where 
\begin{equation}
\tilde{z}_\pm (\nu_j ) = \frac{1}{\sqrt{2}} (\tilde{x} (\nu_j ) \pm i \tilde{y}
(\nu_j ) )~.
\end{equation}
As shown in Ref.[19] the partition function $\mathcal{Z}_S$ in equation (8) can
be written as (cf., also Eq.(25))
\begin{equation}
\mathcal{Z} =\prod_{j=1} ^\infty\mathcal{Z}_j ^+ \mathcal{Z}_j^- ~~,
\end{equation}
where,  
\begin{eqnarray}
\mathcal{Z}_j ^{+} &= &\frac{1}{\sqrt{2\pi\hbar^2 \beta /m}}\int_{-\infty}
^\infty dz_+(0) 
\exp{\left[-\frac{m\beta\omega_0 ^2}{2} |z_+(0)|^2\right] }\nonumber\\
&& \times \prod_{j=1}^\infty \int_{-\infty}^\infty \int_{-\infty}^\infty
\frac{d{\rm Re} z_+ d{\rm Im} z_+}{\pi/(m\beta\nu_j ^2)} \nonumber\\
&\times& \exp\left[-m\beta \left(\nu_j ^2 +\omega_0 ^2 -i\omega_c \nu_j \right)
\left({\rm Re} z_+ ^2 + {\rm Im}z_+^2 \right)\right] ~,\nonumber\\
\end{eqnarray}
and
\begin{equation}
\mathcal{Z}_j ^- =\left(\mathcal{Z}_j ^+ \right)^\ast ~.
\end{equation}
Carrying out the Gaussian integrals we find
\begin{equation}
\mathcal{Z}_j ^+ =\frac{1}{\beta\hbar\omega_0 }~ \frac{\nu_j ^2 }{(\nu_j ^2
+\omega_0 ^2 -i \omega_c \nu_j)}~.
\end{equation}
Hence,
\begin{equation}
\mathcal{Z}_S = \left(\frac{1}{\beta\hbar\omega_0 }\right)^2 \prod_{j=1} ^\infty
\frac{\nu_j ^4 }{\left(\nu_j ^2 + \omega_0 ^2 \right)^2 + \omega_c ^2 \nu_j
^2}~.
\end{equation}
Turning now to the dissipative system described by the full many body
Hamiltonian in Eq.(14) we can similiarly derive \cite{19}
\begin{equation}
\mathcal{Z}(\omega_0) =\frac{1}{(\hbar\beta\omega_{0} )^2 } \prod_{j=1}^{\infty}
\frac{\nu_{j} ^{4} }{(\nu_{j} ^{2} +\omega_{0} ^{2}
+\nu_{j}\tilde{\gamma}(\nu_{j} ))^{2} + \omega_{c} ^{2} \nu_{j} ^{2}}~~,
\end{equation}
where $\gamma(\nu_j)$  is the frequency (ie., $\nu_j$) $-$ dependent friction
coefficient. The Ohmic dissipation model, discussed earlier in Eq.(21) that
yields constant friction, is not suitable for calculating $\mathcal{Z}$ as it
leads to a singularity. In order to regularize the  latter it is convenient to
introduce a `Drude cut-off' by writing the spectral density as (cf., eq.(21))
\begin{equation}
g(\omega)=\frac{2m\gamma}{\pi
\hat{c}^2}.\frac{\omega^2}{1+\frac{\omega^2}{\omega_D ^2}}.
\end{equation}
Correspondingly (cf., Eq.(21)),
\begin{equation}
\tilde{\gamma}(\nu_{j} )=\frac{\gamma\omega_{D} }{(\nu_{j} + \omega_{D}
)}~,~~\nu_j =\frac{2\pi j}{\hbar\beta}~.
\end{equation}
All our results in the sequel are restricted to Ohmic-Drude spectral density
(Eqs.(21) and (44)), though it is known that other forms of frequency-dependence
of the spectral density yield diverse forms of power-law dependence of the
specific heat at low-temperatures \cite{20}.

Inserting this form of the friction coefficient in Eq.(43) the internal energy
$U$ can be calculated as
\begin{eqnarray}
U(\omega_0)&=&-\frac{2}{\beta} - \frac{1}{\beta} \sum_{j=1}^{3} \left[
\frac{\lambda_{j} }{\nu} \psi(\frac{\lambda_{j} }{\nu}) + \frac{\lambda_{j}^{'}
}{\nu} \psi(\frac{\lambda_{j}^{'} }{\nu})\right]\nonumber\\
&+&\frac{2}{\beta}\frac{\omega_D}{\nu}\psi(\frac{\omega_D}{\nu})~,
\end{eqnarray}
where $\psi(z)$ is the digamma function and the arguments are:
\begin{eqnarray}
 \lambda_{1} +\lambda_{2} +\lambda_{3} &=&\omega_{D} +i\omega_{c}\nonumber~,\\
 \lambda_{1} \lambda_{2} +\lambda_{2} \lambda_{3} +\lambda_{3} \lambda_{1}
&=&\omega_{0} ^{2} +\gamma\omega_{D} +i\omega_{c} \omega_{D}\nonumber~,\\
 \lambda_{1} \lambda_{2} \lambda_{3}& =&\omega_{0} ^{2} \omega_{D}\nonumber~.\\
 \end{eqnarray}
The corresponding primed $\lambda$'s are obtained from the complex conjugate of
Eq.(47). Finally, it is easy to derive for the heat capacity the expression
(cf., Eq.(27))\cite{19}
\begin{eqnarray}
C^{\rm Gibbs} _{(\omega_0 \ne 0)} &=&-2k_{B} + k_{B} \sum_{k=1}^{3}
\bigg\{\left(\frac{\lambda_{k}}{\nu}\right)^{2} \psi^{'}
(\frac{\lambda_{k}}{\nu})\nonumber\\
& +& \left(\frac{\lambda_{k}^{'}}{\nu}\right)^{2} \psi^{'}
(\frac{\lambda_{k}^{'}}{\nu})\bigg\} \nonumber\\
&-& 2k_{B}\left(\frac{\omega_{D} }{\nu} \right)^{2} \psi^{'} (\frac{\omega_{D}
}{\nu})~.
\end{eqnarray}
We are now ready to discuss the low and high-temperature limits of the heat
capacity.
\vspace{0.5cm}

({\bf{a}}) \underline{Low- $T$ limit}

\begin{equation}
C^{\rm Gibbs} _{(\omega_0 \ne 0)}=\frac{2\pi}{3} \frac{\gamma}{\omega_0 ^2 }\frac{ k_B ^2 T}{\hbar}+\alpha_1 ^G T^3
+\mathcal{O}(T^5)
\end{equation}
where
\begin{eqnarray*}
 \alpha_1 ^G &=& \frac{8\pi^3}{15}\frac{\gamma}{\omega_0}\frac{k_B ^4}{(\hbar\omega_0)^3}\bigg\{\frac{3(\omega_c ^2 +\omega_0 ^2)}{\omega_0 ^2}\\
&-&(\frac{\gamma}{\omega_0})^2 -\frac{3\omega_0}{\omega_D}(\frac{\gamma}{\omega_0}+\frac{\omega_0}{\omega_D})\bigg\}\\
\end{eqnarray*}

Curiously, to leading order, the presence of the magnetic field through the
cyclotron frequency disappears from $C^{\rm Gibbs} _{(\omega_0 \ne 0)}$, the
expression of which matches with that of a two-dimensional quantum oscillator
(Einstein oscillator). The result in Eq.(49) has been much in discussion in
recent times, in the context of the third law of thermodynamics as it provides a
satisfactory power-law behavior in temperature \cite{4}. 
\vspace{0.5cm}

({\bf{b}})\underline{ Hight -$T$ limit}

At high temperatures ($\hbar \omega_c,~\hbar \omega_0,~\hbar \gamma,~\hbar
\omega_D << k_B T$) our quantum system is expected to be described by classical
statistical mechanics. We find
\begin{equation}
 C^{\rm Gibbs} _{(\omega_0 \ne 0)} =2k_B -\frac{\alpha_2 ^G }{T^2}~.
\end{equation}
where
\begin{eqnarray*}
 \alpha_2 ^G &=& \frac{\hbar^2}{12k_B}(\omega_c ^2 +2\omega_0 ^2 +2\gamma\omega_D)\\
\end{eqnarray*}

In the limit of infinite temperature, therefore, we recover the expected
`equipartition' result:
\begin{equation}
 C^{\rm Gibbs} _{(\omega_0 \ne 0)}=2k_B~,
\end{equation}
where the factor of 2 comes from two dimensions, each of which contributes $k_B$
to the specific heat, $\frac{1}{2}k_B$ arising from the kinetic energy while the
other half from the potential energy.
\subsection{Gibbs Approach ($\omega_0 =0$)}
While studying dissipative Landau diamagnetism we have learnt that taking
$\omega_0 =0$ at the outset yields puzzlingly different result from keeping
$\omega_0$ fixed, evaluating the partition function, calculating its derivatives
and then setting $\omega_0 =0$ \cite{9}. It is already evident from the
low-temperature specific heat (Eq.(49)) that it is not meaningful to take the
limit of $\omega_0 =0$ without `fixing' the coupling with the heat bath
characterized by the friction coefficient $\gamma$ ! It is therefore of 
interest to take a relook at the heat capacity calculation by investigating
afresh the partition function for a charge in a magnetic field  (without the
oscillator potential). In this case only two roots $\lambda_1$ and $\lambda_2$
(cf., Eqs.(46)) matter \cite{19} and we find
\begin{equation}
\mathcal{Z}(\omega_0 =0) = \frac{Nm\beta}{8\pi^{3} } (\gamma^{2} + \omega_{c}
^{2} ) \frac{\prod_{k=1}^{2} \Gamma\left(\frac{\lambda_{k} }{\nu}\right)
\Gamma\left(\frac{\lambda_{k} ^{'} }{\nu}\right)}{\left(\gamma(\frac{\omega_{D}
}{\nu})\right)}~.
\end{equation}
The heat capacity  becomes
\begin{eqnarray}
C^{\rm Gibbs} _{(\omega_0 = 0)}&=&-k_{B} + k_{B} \sum_{k=1}^{2} \bigg\{
\left(\frac{\lambda_{k}}{\nu}\right)^{2} \psi^{'} (\frac{\lambda_{k}}{\nu})
\nonumber\\
&+& \left(\frac{\lambda_{k}^{'}}{\nu}\right)^{2} \psi^{'}
(\frac{\lambda_{k}^{'}}{\nu})\bigg\} \nonumber\\
&-& 2k_{B}\left(\frac{\omega_{D} }{\nu} \right)^{2} \psi^{'} (\frac{\omega_{D}
}{\nu})~.
\end{eqnarray}
We now discuss the low and high temperature limits of Eq.(53).
\vspace{0.5cm}

({\bf{a}}) \underline{Low- $T$ limit}

Using asymptotic expansions as before, we find
\begin{eqnarray}
C^{\rm Gibbs} _{(\omega_0 = 0)}&=&\frac{2\pi}{3} \frac{\gamma}{\hbar}
\frac{(1-\frac{\gamma}{\omega_{D} })}{(\gamma^{2} + \omega_{c} ^{2} )} k_{B}^{2} T \nonumber\\
&-&(\alpha_3 ^G -\alpha_4 ^G)T^3 +O(T^5)~.
\end{eqnarray}
where
\begin{eqnarray*}
 \alpha_3 ^G &=& \frac{8\pi^3}{15}\frac{k_B ^4}{\hbar^3 \sqrt{(\gamma^2 +\omega_c ^2)^3}}\bigg\{\frac{(\gamma^3 -3\gamma\omega_c ^2)}{\sqrt{(\gamma^2 +\omega_c ^2)^3}}(1-\frac{3\gamma}{\omega_D})\\
&+& \frac{(\omega_c ^3 -3\omega_c \gamma^2)}{\sqrt{(\gamma^2 +\omega_c ^2)^3}}\left((\frac{\omega_c}{\omega_D})^3 +3\gamma\frac{\omega_c}{\omega_D ^2}\right)\bigg\}\\
\alpha_4 ^G &=& \frac{8\pi^3}{15}\frac{k_B ^4}{(\hbar\omega_D)^3}\\
\end{eqnarray*}
While Eq.(54) is in conformity with the third law of thermodynamics with
identical linear temperature dependence as in the case of $\omega_0 \ne 0$, but,
is free from the singularity issue in Eq.(49) (for $\omega_0 =0$). It leads, in
the limit of  $\omega_D =\infty$ (infinite Drude cut-off) to the result:
\begin{equation}
 C^{\rm Gibbs} _{(\omega_0 = 0)}=\frac{2\pi}{3\hbar}k_B ^2 T
\frac{\gamma}{\gamma^2 +\omega_c ^2 }~.
\end{equation}
Further, for very strong magnetic fields ($\gamma << \omega_c$), 
\begin{equation}
 C^{\rm Gibbs} _{(\omega_0 = 0)}=\frac{2\pi}{3}\frac{\gamma}{\omega_c
^2}\frac{k_B ^2 T}{\hbar}~,
\end{equation}
a harmonic oscillator like result with the cyclotron frequency $\omega_c$
replacing $\omega_0$. On the other hand, for weak magnetic fields ($\gamma >>
\omega_c$),
\begin{equation}
 C^{\rm Gibbs} _{(\omega_0 = 0)}=\frac{2\pi}{3}\frac{k_B ^2
T}{\hbar}\frac{1}{\gamma}~,
\end{equation}
the free particle result in which the friction coefficient $\gamma$ appears in
the denominator, in agreement to the corresponding result given in \cite{21},
after a proper counting of the degree of freedom. 
\vspace{0.5cm}

({\bf{b}})\underline{ Hight -$T$ limit}

We find
\begin{equation}
 C^{\rm Gibbs} _{(\omega_0 = 0)} =k_B -\frac{\hbar^2}{12k_B T^2}(\omega_c ^2
+2\gamma\omega_D)
\end{equation}
Again, equipartition theorem for a free particle (in 2 dimensions) prevails at
$T=\infty$. 

Thus the classical limit of the Landau problem, as far as the heat capacity is
concerned, is that of free particle whereas an additional (parabolic)
constraining potential yields harmonic oscillator behavior.
\subsection{Einstein approach ($\omega_0 \ne 0$)}
We will now focus on the Einstein approach based on the Langevin equation (15)
which can be recast into the following convenient form \cite{9}:
\begin{equation}
 \ddot{z}+\int_0 ^t dt^{\prime}
\bar{\gamma}(t-t^{\prime})\dot{z}(t^{\prime})+\omega_0 ^2 z =\frac{F(t)}{m}~,
\end{equation}
where 
\begin{equation}
z=x+iy, ~~F=f_x +if_y, \rm{and} ~ \bar{\gamma}(t)=\gamma(t)+i\omega_c~.
\end{equation}
 In order to find the time-dependent specific heat we need the internal energy
which is the statistical average of the Hamiltonian given by
\begin{equation}
 \mathcal{H}=\frac{1}{2}m\dot{z}\dot{z}^{\dagger}-\frac{1}{2}\hbar\omega_c
+\frac{1}{2}m\omega_0 ^2 zz^{\dagger}.
\end{equation}
We therefore need the equal-time correlation functions:
\begin{subequations}
\begin{align}
 \zeta_1 (t)=\langle z(t)z^{\dagger}(t) \rangle ~,\\
\zeta_2 (t)=\langle \dot{z}(t)\dot{z}^{\dagger}(t) \rangle~.
\end{align}
\end{subequations}
The correlation functions in Eq.(62) can be found from the analytic continuation
to $t^{\prime}=t$ of the unequal time correlation functions, eg.,
\begin{equation}
 \zeta_1 (t,t^{\prime})=\langle z(t)z^{\dagger}(t^{\prime}) \rangle ~,
\end{equation}
where $z(t)$ can be further expressed in terms of the response function
$\chi(t)$ as 
\begin{equation}
 z(t)=\int_0 ^t d\tau \chi(t-\tau)\frac{F(\tau)}{m}~.
\end{equation}
The former is the inverse Fourier transform of $\chi(\omega)$ that can be easily
written from Eq.(59) as
\begin{equation}
 \chi (\omega)=\frac{1}{2\pi}\frac{1}{(-\omega^2 -i\omega \bar{\gamma}+\omega_0
^2)}~,
\end{equation} with
\begin{equation}
 \bar{\gamma}(\omega)=i\omega_c +\gamma(\omega)=i\omega_c +\gamma
\frac{\omega_D}{\omega_D -i\omega}~.
\end{equation}
From Eq.(63),
\begin{eqnarray}
\zeta_1 (t, t^{\prime}) &=& \int_0 ^t d\tau \int_0 ^{t^{\prime}} d\tau^{\prime}
\chi(t-\tau)\chi^* (t^{\prime} -\tau^{\prime})\nonumber\\
&\times& \frac{\langle F(\tau)F^{\dagger}(\tau^{\prime})\rangle}{m}~,
\end{eqnarray}
where \cite{9},
\begin{equation}
 \langle F(\tau)F^{\dagger}(\tau^{\prime}) \rangle =\int_{-\infty}^{+\infty}
d\tilde{\omega}f(\tilde{\omega})e^{-i\omega(\tau-\tau^{\prime})}~,
\end{equation}
with
\begin{equation}
 f(\tilde{\omega})=\frac{m}{\pi} \frac{\gamma \omega_D ^2}{(\omega_D ^2
+\tilde{\omega}^2)}\hbar\tilde{\omega}[\coth(\frac{\hbar\tilde{\omega}}{2kT})-1]
~.
\end{equation}

Our strategy is to first calculate $\zeta_1 (t, t^{\prime})$ and $\zeta_2 (t,
t^{\prime})$ (for details, see the Appendix A), then set $t=t^{\prime}$ and
finally, in order to extract the thermal equilibrium internal energy $E$, take
the limit $t=\infty$. We find 
\begin{eqnarray}
 E&=&\langle \mathcal{H} \rangle =-\frac{1}{2}\hbar\omega_c
+\frac{1}{2}m\omega_0 ^2 \lim_{t\rightarrow\infty}\zeta_1 (t)+\frac{1}{2}m
\lim_{t\rightarrow\infty}\zeta_2 (t)\nonumber\\
&=&2k_B T+\frac{\hbar}{2\pi}\sum_{j=1}^3
\bigg\{\psi(1+\frac{\lambda_j}{\nu})[2\omega_0 ^2 q_j +p_j]\nonumber\\
&+&\psi(1+\frac{\lambda^{\prime}_j}{\nu})[2\omega_0 ^2 q^{\prime}_j
+p^{\prime}_j]\bigg\}~,
\end{eqnarray}
where
\begin{subequations}
\begin{align}
 q_j = \frac{(\lambda_j -\omega_D)}{\prod_{j^{\prime}} ^{\prime}(\lambda_j
-\lambda_j^{\prime})}~,\\
p_j =\frac{\lambda_j [\gamma\omega_D -i\omega_c (\lambda_j -\omega_D)]
}{\prod_{j^{\prime}} ^{\prime}(\lambda_j -\lambda_{j^{\prime}})}~.
\end{align}
\end{subequations}
In the denominators of Eqs.(71), the notation $\prod_{j^{\prime}} ^{\prime}$ implies that
the $j=j^{\prime}$ terms are excluded from the product. The quantities
$q^{\prime}_j$ and $p^{\prime}_j$ are obtained by priming the $\lambda'{\rm
{s}}$, the latter having been already defined in Eq.(47).

Finally, the equilibrium specific heat is given by
\begin{eqnarray}
 C_{(\omega_0 \ne 0)} ^{Einstein} &=& \frac{\partial E}{\partial T}\nonumber\\
&=&-2k_B -k_B \beta \frac{\hbar}{2\pi} \sum_{j=1}^3
\bigg\{\frac{\lambda_j}{\nu}\psi^{\prime}(\frac{\lambda_j}{\nu})[2\omega_0 ^2
q_j +p_j]\nonumber\\
&+&\frac{\lambda^{\prime}_j}{\nu}\psi^{\prime}(\frac{\lambda^{\prime}_j}{\nu})[
2\omega_0 ^2 q^{\prime}_j +p^{\prime}_j]\bigg\}~,
\end{eqnarray}
where $\psi^{\prime}(z)$ are the trigamma functions \cite{19}.

We may now discuss the low and the high temperature limits of Eq.(72).
\vspace{0.5cm}

(a)\underline{Low-T limit}
\vspace{0.5cm}

Employing the asymptotic expansion of the digamma function:
\begin{equation}
 \psi^{\prime}(z)=\frac{1}{z}+\frac{1}{2z^2}+\frac{1}{6z^3}-\frac{1}{30z^5}
-....~,
\end{equation}
we find
\begin{equation}
 C^{Einstein} _{\omega\ne0}=\frac{2\pi}{3}\frac{\gamma}{\omega_0 ^2}\frac{k_B
^2 T}{\hbar}+\alpha_1 ^E T^3 +O(T^5)~.
\end{equation}
where
\begin{eqnarray*}
 \alpha_1 ^E &=& \frac{8\pi^3}{15}\frac{\gamma}{\omega_0}\frac{k_B ^4}{(\hbar\omega_0)^3}\bigg\{\frac{3(\omega_c ^2 +\omega_0 ^2)}{\omega_0 ^2}\\
&-&(\frac{\gamma}{\omega_0})^2 -\frac{\omega_0}{\omega_D}(\frac{\omega_c ^2}{\gamma\omega_0}-\frac{2\gamma}{\omega_0}-\frac{\omega_0}{\omega_D})\bigg\}\\
\end{eqnarray*}
As required by the third law of thermodynamics the specific heat does vanish as
a power law as $T\rightarrow0$, exactly in the same manner as in the
corresponding Gibbs expression (cf., Eq.(49)), but interestingly the coefficient
of the next higher order term $(\propto T^3)$ differs from the Gibbs result.
\vspace{0.5cm}

(b)\underline{High-T limit}
\vspace{0.5cm}

At high temperatures,
\begin{equation}
 C^{Einstein} _{\omega\ne0}=2k_B-\frac{\alpha_2 ^E }{T^2}~.
\end{equation}
where
\begin{eqnarray*}
 \alpha_2 ^E = \frac{\hbar^2}{12k_B}(\omega_c ^2 +2\omega_0 ^2 +\gamma\omega_D)
\end{eqnarray*}
At infinite temperatures the classical equipartion result is restored. But again,
in the next higher order term (in $\frac{1}{T^2}$),  the Einstein result differs
from the Gibbs result by a cut-off-dependent term:
\begin{eqnarray}
 C^{Einstein} _{\omega\ne0}=C^{Gibbs} _{\omega\ne0}+\frac{\hbar^2
\gamma\omega_D}{12k_B T^2}~.
\end{eqnarray}
\subsection{Einstein approach ($\omega_0 =0$)}
We now return to discuss the Einstein result for the specific heat due to the
presence of the magnetic field alone, ie., in the absence of the parabolic well.
The relevant Hamiltonian is 
\begin{equation}
 \mathcal{H}=-\frac{1}{2}\hbar\omega_c +\frac{1}{2}m\dot{z}\dot{z}^{\dagger}~,
\end{equation}
and hence
\begin{equation}
 E=-\frac{1}{2}\hbar\omega_c +\frac{1}{2}m\lim_{t\rightarrow\infty}[\zeta_2
(t)]_{\omega_0 =0}~.
\end{equation}
As discussed in Ref.[19], one of the three roots, viz. $\lambda_1$ vanishes for
$\omega_0 =0$. Consequently (see Appendix B, for details), 
\begin{eqnarray}
 E(\omega_0 =0)&=&k_B T \nonumber\\
&+&\frac{\hbar}{2\pi}\{p_2 \psi(1+\frac{\lambda_2}{\nu})+p_3
\psi(1+\frac{\lambda_3}{\nu})\nonumber\\
&+&p^{\prime}_2 \psi(1+\frac{\lambda^{\prime}_2}{\nu})+p^{\prime}_3
\psi(1+\frac{\lambda^{\prime}_3}{\nu})\}~.
\end{eqnarray}
As before, the derivative of $E$ with respect to temperature yields an
expression for the specific heat in terms of the digamma functions, which can be
further analyzed in the low- and high-temperature limits.
\vspace{0.5cm}

(a)\underline{Low-T limit}
\vspace{0.5cm}

Again, using the asymptotic expansion of the digamma function (cf., Eq.(73)), we
find
\begin{equation}
 C^{Einstein} _{\omega=0}=\frac{2\pi}{3}\frac{\gamma}{\hbar}\frac{1}{\gamma^2
+\omega_c ^2} k_B ^2 T -\alpha_3 ^E T^3 +O(T^5)~.
\end{equation}
where
\begin{eqnarray*}
\alpha_3 ^E &=& \frac{8\pi^3}{15}\frac{k_B ^4}{\hbar^3 \sqrt{(\gamma^2 +\omega_c ^2)^3}}\bigg\{\frac{(\gamma^3 -3\gamma\omega_c ^2)}{\sqrt{(\gamma^2 +\omega_c ^2)^3}}\\
&\times&\left(1-\frac{2\gamma}{\omega_D}-(\frac{\omega_c }{\omega_D})^2\right)\\
&+& 10(\frac{\omega_c}{\omega_D})^2 \frac{\gamma(\gamma^2 +\omega_c ^2)}{\sqrt{(\gamma^2 +\omega_c ^2)^3}}\bigg\}\\
\end{eqnarray*}
While the expression in Eq.(80) is in conformity with the third law of
thermodynamics, as expected, it differs from the corresponding Gibbsian result
of Eq.(54) in terms of different dependencies on the Drude cut-off $\omega_D$!
Apart from this issue the strong and weak magnetic field cases follow the
behavior discussed earlier, below Eq.(54).
\vspace{0.5cm}

(b)\underline{High-T limit}
\vspace{0.5cm}
\begin{equation}
 C^{Einstein} _{\omega_0 =0}=k_B -\frac{\hbar^2}{12k_B T^2}(\omega_c ^2 +\gamma\omega_D)~.
\end{equation}
Finally, in the high-temperature limit, equipartion result obtains, but once
again, there is a correction term over and above the Gibbs result that is
cut-off dependent, as we found earlier in the $\omega_0 \ne 0$ case in Eq.(76): 
\begin{eqnarray}
 C^{Einstein} _{\omega_0 =0}=C^{Gibbs} _{\omega_0 =0}+\frac{\hbar^2
\gamma\omega_D}{12k_B T^2}~,
\end{eqnarray}
where $C^{Gibbs} _{\omega_0 =0}$ is given by the high-T expression in Eq.(58).
\section{Summary}
Summarising, we study the various limiting behavior of the specific heat of a
dissipative charged harmonic oscillator in a uniform magnetic field, obtained
from the partition function approach (Gibbs' method) and from the steady state of
corresponding quantum Langevin equation (Einstein's approach). The specific heat
obtained from both these methods shows linear $T$ dependence at low temperatures,
which is in agreement with the third law of thermodynamics. At high temperatures
the specific heat approaches a constant value depending on the number of
degrees of freedom of the system. Although, both the Gibbs and Einstein
approaches are in conformity with the third law of thermodynamics and the
equipartiton theorem, at low and high temperatures respectively, they differ
from each other in detail, beyond the leading order. In the limit of vanishing
confinement frequency ($\omega_0 \rightarrow0$), the specific heat of the oscillator
becomes singular at low-temperatures and manifests extra degrees of freedom counting
at high temperatures. The specific heat of the free particle cannot be obtained from
the equilibrium value ($t\rightarrow \infty$) of the specific heat of the
oscillator just by taking the $\omega_0 \rightarrow0$ limit. It is evident that the
order in which one takes the $t=\infty$ and $\omega_0 =0$ limits yield
qualitatively different answers for the specific heat. While in the Einstein
approach, the free particle-like specific heat emerges by taking the $\omega_0
=0$ limit first before considering the $t=\infty$ limit, the Gibbs approach is
plagued by a singularity issue, for $\omega_0 =0$, in the low-temperature limit
(cf., Eq.(49)). 

\begin{widetext}
\begin{table*}[t]
\small
  \begin{tabular}{|l| l| l|l|l|l|l|}
\hline
    & \multicolumn{2}{|c|}{$\omega_c \neq 0$, $\omega_0 \neq 0$} & \multicolumn{2}{|c|}{$\omega_c \neq 0$, $\omega_0 = 0$} & \multicolumn{2}{|c|}{$\omega_c = 0$, $\omega_0 = 0$} \\
\hline
 & Low  & High  & Low  & High  & Low  & High  \\
& Temperature & Temperature & Temperature & Temperature & Temperature & Temperature \\
\hline
&&&&&&\\
Gibbs  &$\frac{2\pi}{3}\frac{\gamma}{\omega_0 ^2}\frac{k_B ^2 T}{\hbar}$  & $2k_B-\frac{\alpha_2 ^G}{T^2}$ &  $\frac{2\pi}{3}\frac{\gamma}{\hbar}\frac{(1-\frac{\gamma}{\omega_D})}{\gamma^2 +\omega_c ^2}k_B ^2 T$ &$k_B-\frac{\alpha_2 ^G |_{\omega_0=0}}{T^2}$  &$\frac{2\pi}{3}\frac{k_B ^2 T}{\hbar\gamma}(1-\frac{\gamma}{\omega_D})$  & $k_B -\frac{\alpha_2 ^G |_{(\omega_0 =0,~\omega_c =0)}}{T^2}$ \\
Approach& ${\small -\alpha_1 ^G T^3 +O(T^5)}$ & ~ & $-(\alpha_3 ^G -\alpha_4 ^G)T^3 $ & ~&$-(\alpha_3 ^G |_{\omega_c =0} -\alpha_4 ^G)T^3$ &~\\
& ~ & ~ & $+O(T^5)$ & ~ &$+O(T^5)$ &~\\
&&&&&&\\
\hline
&&&&&&\\
Einstein  & $\frac{2\pi}{3}\frac{\gamma}{\omega_0 ^2}\frac{k_B ^2 T}{\hbar}$  &$2k_B -\frac{\alpha_2 ^E}{T^2}$ & $\frac{2\pi}{3}\frac{\gamma}{\hbar}\frac{1}{\gamma^2 +\omega_c ^2}k_B ^2 T$ & $k_B -\frac{\alpha_2 ^E |_{\omega_0 =0}}{T^2}$ &$\frac{2\pi}{3}\frac{k_B ^2 T}{\hbar\gamma}-\alpha_3 ^E |_{\omega_c =0} T^3$  & $k_B -\frac{\alpha_2 ^E |{(\omega_c =0,~\omega_0 =0)}}{T^2}$ \\
Approach& $-\alpha_1 ^E T^3 +O(T^5)$ & ~ & $-\alpha_3 ^E T^3 +O(T^5)$ &~ &$+O(T^5)$ &~\\
&&&&&&\\
\hline
  \end{tabular}
\label{t1}
\caption{ Comparison of Specific Heat in the Gibbs Approach and the Einstein Approach in different limits.}
\end{table*}
\begin{table*}
\small
\begin{tabular}{|l| l| l|l|}
\hline
& \multicolumn{2}{|c|}{Specific Heat}&Magnetization\\
\hline
&Low Temperature&High Temperature&($\gamma\rightarrow0$)~~~~~~~\\
\hline
&&&\\
$\omega_0\rightarrow0,t\rightarrow\infty$&$\frac{2\pi}{3}\frac{\gamma}{\hbar}\frac{1}{\gamma^2 +\omega_c ^2}k_B ^2 T -\alpha_3 ^E T^3 +O(T^5)$ &$k_B -\frac{\alpha_2 ^E |{\omega_0 =0}}{T^2}$&$-\frac{|e|\hbar}{2mc}\coth(\frac{\hbar\omega_c}{2k_B T})$\\
&&&\\
\hline
&&&\\
$t\rightarrow\infty,\omega_0\rightarrow0$&Singularity&$2k_B -\frac{\alpha_2 ^E |_{\omega_0 =0}}{T^2}$&$\frac{|e|\hbar}{2mc}[\frac{2k_B T}{\hbar\omega_c}-\coth(\frac{\hbar\omega_c}{2k_B T})]$\\ 
&&&\\
\hline
\end{tabular}
\caption{Specific Heat and Magnetization in the limit of vanishing confinement frequency in two sequences.}
 \label{t2}
\end{table*}
\end{widetext}
In Table I, we summarise our results for the Specific Heat in different limits for both the Gibbs and Einstein approaches. In the limit of $\omega_D\rightarrow \infty$, both the Gibbs and Einstein approaches give the same thermodynamic results. However, for a finite cutoff frequency $\omega_D$, the results differ in next to the leading order at both high and low temperatures.
The results summarized in Table I lead to the following conclusions :
\begin{enumerate}
 \item At low temperatures the specific heat is linear in temperature and hence the dissipative environment restores the third law of thermodynamics. 
\item In the presence of the oscillator potential, the low temperature behavior of the specific heat goes as $1/\omega_0^2$ and is therefore singular in the limit of $\omega_0\rightarrow 0$. Thus the results of the unconfined particle cannot be recovered in this limit.
\item The high temperature specific heat approaches a constant value independent of the confinement potential and depends only on the number of degrees of freedom in agreement with the equipartition law. Again, the results of the unconfined system cannot be recovered in the limit of vanishing confinement frequency $\omega_0$. 
\end{enumerate}
While the issue of recovering the results of the unconfined particle, starting from the confined system and taking the limit of vanishing confinement frequency $\omega_0$ cannot be resolved at the equilibrium level, the Einstein approach  has the intrinsic advantage of obtaining the results in the process of equilibration. The equilibrium results can be arrived at by taking the limit of $t\rightarrow \infty$. Hence, one could in principle ask the question, what would happen if the confinement frequency $\omega_0$ is taken to zero, before the limit $t \rightarrow \infty$ is taken. A similiar result was obtained for the case of a particle in a harmonic oscillator potential\cite{22}. The results for the two different sequences of taking the limits is summarised in Table \ref{t2}. It is clear from the table that, if the limit of $\omega_0 \rightarrow 0$ is taken before the limit of $t \rightarrow \infty$, one can actually recover the results of the unconfined system for the specific heat and magnetization. 
It is curious to note that the result for magnetization obtained from this sequence of taking the limits is inconsistent with the Landau results, whereas when the limits are taken in the other way round, the Landau result is recovered. This is, however, due to the fact that the Landau result for magnetization can only be recovered in the presence of a confinement potential. 
\section{Acknowledgements}
We thank Malay Bandyopadhyay and Gert Ingold for useful discussions. SD is
grateful to the J. C. Bose Fellowship of the Department of Science and
Technology for supporting this work.
\appendix
\vskip 1cm
\section{Einstein Approach ($\omega_0 \ne 0$)}
With the help of the Drude cut-off frequency we can write $\chi(\omega)$ as
\begin{small}
\begin{equation}
\chi(\omega)=\frac{(\omega_D -i\omega)}{[i\omega^3 -\omega^2 (\omega_D
+i\omega_c)-i\omega(\gamma\omega_D +i\omega_c \omega_D +\omega_0 ^2)+\omega_0 ^2
\omega_D]}
\end{equation}
\end{small}
Alternatively,
\begin{equation}
 \chi(\omega)=-\frac{(\omega+i\omega_D)}{
(\omega+i\lambda_1)(\omega+i\lambda_2)(\omega+i\lambda_3)}~,
\end{equation}
where $\lambda_j s$ and $\lambda^{\prime}_j s$ are given by the Vieta equations
(Eq.(47)). We can write Eq.(67) as
\begin{eqnarray}
 \zeta_1 (t,t^{\prime})&=&\langle z(t)z^{\dagger}(t^{\prime})\rangle
=\frac{1}{4\pi^2 m^2}
\int_{-\infty}^{+\infty}d\tilde{\omega}f(\tilde{\omega})\nonumber\\
&\times&\int_{-\infty}^{+\infty}d\omega
\chi(\omega)\frac{(e^{-i\tilde{\omega}t}-e^{-i\omega
t})}{i(\omega-\tilde{\omega})}\nonumber\\
&\times&\int_{-\infty}^{+\infty}d\omega^{\prime}\chi^*
(\omega^{\prime})\frac{(e^{i\tilde{\omega}t^{\prime}}-e^{i\omega^{\prime}
t^{\prime}})}{-i(\omega^{\prime}-\tilde{\omega})}~.
\end{eqnarray}
The two integrals, defined by
\begin{eqnarray}
I_1 &=&\int_{-\infty}^{+\infty}d\omega
\chi(\omega)\frac{(e^{-i\tilde{\omega}t}-e^{-i\omega
t})}{i(\omega-\tilde{\omega})}~,\\
I_2 &=&\int_{-\infty}^{+\infty}d\omega^{\prime}\chi^*
(\omega^{\prime})\frac{(e^{i\tilde{\omega}t^{\prime}}-e^{i\omega^{\prime}
t^{\prime}})}{-i(\omega^{\prime}-\tilde{\omega})}~,
\end{eqnarray}
can be expressed as 
\begin{eqnarray}
I_1 &=& \frac{2\pi}{iA}\{\frac{(\lambda_1 -\omega_D)(\lambda_2
-\lambda_3)(e^{-i\tilde{\omega}t}-e^{-\lambda_1
t})}{(\tilde{\omega}+i\lambda_1)}\nonumber\\
&+&\frac{(\lambda_2 -\omega_D)(\lambda_3
-\lambda_1)(e^{-i\tilde{\omega}t}-e^{-\lambda_2
t})}{(\tilde{\omega}+i\lambda_2)}\nonumber\\
&+&\frac{(\lambda_3 -\omega_D)(\lambda_1
-\lambda_2)(e^{-i\tilde{\omega}t}-e^{-\lambda_3
t})}{(\tilde{\omega}+i\lambda_3)}\}~,\\
I_2 &=& -\frac{2\pi}{i A^{\prime} }\{ \frac{(\lambda^{\prime} _{1}
-\omega_D)(\lambda^{\prime}_2
-\lambda^{\prime}_3)(e^{i\tilde{\omega}t^{\prime}}-e^{-\lambda^{\prime}_1
t^{\prime}})}{(\tilde{\omega}-i\lambda^{\prime}_1)}\nonumber\\
&+&\frac{(\lambda^{\prime}_2 -\omega_D)(\lambda^{\prime}_3
-\lambda^{\prime}_1)(e^{i\tilde{\omega}t^{\prime}}-e^{-\lambda^{\prime}_2
t^{\prime}})}{(\tilde{\omega}-i\lambda^{\prime}_2)}\nonumber\\
&+&\frac{(\lambda^{\prime}_3 -\omega_D)(\lambda^{\prime}_1
-\lambda^{\prime}_2)(e^{i\tilde{\omega}t^{\prime}}-e^{-\lambda^{\prime}_3
t^{\prime}})}{(\tilde{\omega}-i\lambda^{\prime}_3)}\}~.
\end{eqnarray}
where
\begin{eqnarray}
 A&=&(\lambda_1 -\lambda_2)(\lambda_1 -\lambda_3)(\lambda_2 -\lambda_3)~,\\
A^{\prime}&=&(\lambda^{\prime}_1 -\lambda^{\prime} _2)(\lambda^{\prime}_1
-\lambda^{\prime}_3)(\lambda^{\prime}_2 -\lambda^{\prime}_3)~.
\end{eqnarray}
Eq.(A3) then yields
\begin{eqnarray}
\zeta_1 (t, t^{\prime})&=&\frac{1}{4\pi^2 m^2}\int_{-\infty}^{+\infty}
d\tilde{\omega}f(\tilde{\omega})I_1 I_2 \nonumber\\
&=&\frac{1}{4\pi^2 m^2}\int_{-\infty}^{+\infty} d\tilde{\omega}\frac{m}{\pi}
\frac{\gamma \omega_D ^2}{(\omega_D ^2
+\tilde{\omega}^2)}\hbar\tilde{\omega}\nonumber\\
&\times&\coth(\frac{\hbar\tilde{\omega}}{2kT})I_1 I_2 \nonumber\\
&-&\frac{1}{4\pi^2 m^2}\int_{-\infty}^{+\infty} d\tilde{\omega}\frac{m}{\pi}
\frac{\gamma \omega_D ^2}{(\omega_D ^2 +\tilde{\omega}^2)}\hbar\tilde{\omega}I_1
I_2~.\nonumber\\
\end{eqnarray}
The second integral vanishes for symmetry reasons, so that only the integral
containing cotangent hyperbolic contributes.  In order to find out the equal
time correlation function $\zeta_1 (t)$, we set $t=t^{\prime}$. In that case 
the coefficients of $e^{-i\tilde{\omega}t}$ and $e^{i\tilde{\omega}t^{\prime}}$
matter,  because in the product, these are the only time independent parts. Now
substituting Eqs.(A6) and (A7) in (A10), we can easily separate the mean squared
average into two parts, one that is completely time independent and the other
which is an exponentially decaying (time dependent) one. In the limit of
$t\rightarrow\infty$, the time dependent parts vanish and we are left with the
equilibrium value. Finally,
\begin{eqnarray}
 \zeta_1 (t)&=&\frac{\hbar}{4\pi^3 m}\int_{-\infty}^{+\infty}
d\tilde{\omega}\frac{\gamma \omega_D ^2}{(\omega_D ^2
+\tilde{\omega}^2)}\tilde{\omega}\nonumber\\
&\times& \coth(\frac{\hbar\tilde{\omega}}{2k_B T})I^{\prime}_1 I^{\prime}_2 ~.
\end{eqnarray}
where
\begin{eqnarray}
I^{\prime}_1 &=& \frac{2\pi}{i}\frac{(\omega_D
-i\tilde{\omega})e^{-i\tilde{\omega}t}}{(\tilde{\omega}+i\lambda_1)(\tilde{
\omega}+i\lambda_2)(\tilde{\omega}+i\lambda_3)}~,\\
I^{\prime}_2 &=& -\frac{2\pi}{i}\frac{(\omega_D
+i\tilde{\omega})e^{i\tilde{\omega}t}}{(\tilde{\omega}-i\lambda^{\prime}
_1)(\tilde{\omega}-i\lambda^{\prime}_2)(\tilde{\omega}-i\lambda^{\prime}_3)}~.
\end{eqnarray}
We can write
\begin{equation}
\zeta_1 (t)=\langle z(t)z^{\dagger}(t) \rangle =Q_1 -Q_2~.
\end{equation}
where
\begin{eqnarray}
 Q_1 &=&-\frac{\hbar}{2\pi m}\int_{-\infty}^{+\infty}
d\tilde{\omega}\coth(\frac{\hbar\tilde{\omega}}{2kT})\nonumber\\
&\times&\frac{(\omega_D
-i\tilde{\omega})}{(\tilde{\omega}+i\lambda_1)(\tilde{\omega}+i\lambda_2)(\tilde
{\omega}+i\lambda_3)}~,\nonumber\\
Q_2 &=& \frac{\hbar}{2 \pi
m}\int_{-\infty}^{+\infty}d\tilde{\omega}\coth(\frac{\hbar\tilde{\omega}}{2kT})
\nonumber\\
&\times&\frac{(\omega_D
+i\tilde{\omega})}{(\tilde{\omega}-i\lambda^{\prime}_1)(\tilde{\omega}-i\lambda^
{\prime}_2)(\tilde{\omega}-i\lambda^{\prime}_3)}~.\nonumber\\
\end{eqnarray}
Assuming that the time is long enough compared to the relaxation time, we can
ignore the integrals containing $\lambda_1$, $\lambda_2$, $\lambda_3$.  After
simplifications
\begin{eqnarray}
\zeta_1 (t)&=& \langle z(t)z^{\dagger}(t)\rangle=\frac{2kT}{m\omega_0
^2}\nonumber\\
&+&\frac{\hbar}{m\pi}\sum_{j=1}^3 \bigg\{ q_j
\psi(1+\frac{\lambda_j}{\nu})+q^{\prime} _j
\psi(1+\frac{\lambda^{\prime}_j}{\nu})\bigg\}~.\nonumber\\
\end{eqnarray}
where $\psi(1+z_j)$ is a digamma function, $\nu =\frac{2\pi kT}{\hbar}$, and the
$q_{j}$ and the $q^{\prime}_{j}$ are defined in Eq.(71a). We can observe from
Eq.(A16),(since $\langle \vec{r}^2 \rangle =\langle z(t)z^{\dagger}(t)\rangle $)
that the equipartition theorem is satisfied for this two-dimensional problem.

We will calculate $\zeta_2 (t,t^{\prime})$, which is defined as
\begin{eqnarray}
 \zeta_2 (t,t^{\prime})&=&\langle \dot{z}(t)\dot{z}^{\dagger}(t)\rangle
=\frac{\hbar}{m\pi}\int_{-\infty}^{+\infty}d\omega \omega^2
\chi^{\prime\prime}\coth(\frac{\hbar\omega}{2kT})\nonumber\\
&-&\frac{\hbar}{m\pi}\int_{-\infty}^{+\infty}d\omega \omega^2
\chi^{\prime\prime}\\
&=&\frac{2kT}{m}+\frac{\hbar\omega_0 ^2}{m\pi}\sum_{j=1}^3 \bigg\{ q_j
\psi(1+\frac{\lambda_j}{\nu})+q^{\prime}_j
\psi(1+\frac{\lambda^{\prime}_j}{\nu})\bigg\}\nonumber\\
&-&\frac{\hbar}{m\pi}\bigg\{ \sum_{j=1}^3 p_j \sum_{n=1}^\infty
\frac{1}{n+\frac{\lambda_j}{\nu}}+\sum_{j=1}^3 p^{\prime}_j \sum_{n=1}^\infty
\frac{1}{n+\frac{\lambda^{\prime}_j}{\nu}}\bigg\}\nonumber\\
&+&\frac{\hbar\omega_c}{m}~,
\end{eqnarray}
where $q_{j}$ and the $q^{\prime}_{j}$ are defined in Eq.(71a), and  $p_j$ and
$p^{\prime}_j $ are given by (71b). 
We now use a transformation $P_j =p_j +i\frac{\omega_c}{3}$,  such that
$\sum_{j=1}^3 P_j=0$, since $\sum_{j=1}^3 p_j =-i\omega_c$. Therefore
\begin{eqnarray}
 \sum_{j=1}^3 p_j \sum_{n=1}^\infty
\frac{1}{n+\frac{\lambda_j}{\nu}}&=&\sum_{j=1}^3 P_j \sum_{n=1}^\infty
\frac{1}{n+\frac{\lambda_j}{\nu}}\nonumber\\
&-&\sum_{j=1}^3 i\frac{\omega_c}{3}\sum_{n=1}^\infty
\frac{1}{n+\frac{\lambda_j}{\nu}}\nonumber\\
&=&-\sum_{j=1}^3 P_j \psi(1+\frac{\lambda_j}{\nu})\nonumber\\
&-&\sum_{j=1}^3 i\frac{\omega_c}{3} \sum_{n=1}^\infty
\frac{1}{n+\frac{\lambda_j}{\nu}}~.
\end{eqnarray}
In a similiar fashion we can use a transformation $P^{\prime}_j =p^{\prime}_j
-i\frac{\omega_c}{3}$,  in such a way that $\sum_{j=1}^3 P^{\prime}_j =0$ since
$\sum_{j=1}^3 p^{\prime}_j =i\omega_c$, hence
\begin{eqnarray}
 \sum_{j=1}^3 p^{\prime}_j \sum_{n=1}^\infty
\frac{1}{n+\frac{\lambda^{\prime}_j}{\nu}}&=&\sum_{j=1}^3 P^{\prime}_j
\sum_{n=1}^\infty \frac{1}{n+\frac{\lambda^{\prime}_j}{\nu}}\nonumber\\
&+&\sum_{j=1}^3 i\frac{\omega_c}{3}\sum_{n=1}^\infty
\frac{1}{n+\frac{\lambda^{\prime}_j}{\nu}}\nonumber\\
&=&-\sum_{j=1}^3 P^{\prime}_j \psi(1+\frac{\lambda^{\prime}_j}{\nu})\nonumber\\
&+&\sum_{j=1}^3 i\frac{\omega_c}{3} \sum_{n=1}^\infty
\frac{1}{n+\frac{\lambda^{\prime}_j}{\nu}}~.
\end{eqnarray}
 Substituting Eqs.(A19) and (A20) in Eq.(A18) and using three important
properties of the digamma functions \cite{23}
\begin{eqnarray}
 \psi(x)-\psi(y)&=&\frac{(x-y)}{xy}+\sum_{n=1}^\infty
[\frac{1}{n+y}-\frac{1}{n+x}]~,\nonumber\\
\psi(1+z)&=&\psi(z)+\frac{1}{z}~,\nonumber\\
\sum_{j=1}^N a_j \sum_{n=1}^\infty \frac{1}{n+z_j}&=&-\sum_{j=1}^N a_j
\psi(1+z_j)~~~[\sum_{j=1}^N a_j =0]~,\nonumber\\
\end{eqnarray}
we obtain 
\begin{eqnarray}
\zeta_2 (t)&=& \langle \dot{z}(t)\dot{z}^{\dagger}(t) \rangle =
\frac{2kT}{m}\nonumber\\
&+&\frac{\hbar\omega_0 ^2}{m\pi}\sum_{j=1}^3 \bigg\{q_j
\psi(1+\frac{\lambda_j}{\nu})+q^{\prime}_j
\psi(1+\frac{\lambda^{\prime}_j}{\nu})\bigg\}\nonumber\\
&+&\frac{\hbar}{m\pi}\sum_{j=1}^3 \bigg\{p_j
\psi(1+\frac{\lambda_j}{\nu})+p^{\prime}_j
\psi(1+\frac{\lambda^{\prime}_j}{\nu})\bigg\}\nonumber\\
&+&\frac{\hbar\omega_c}{m}~.
\end{eqnarray}
From Eq.(A22), we can calculate the mean squared average of the kinematic
momentum of the particle in a magnetic field, given by
\begin{eqnarray}
\langle (\vec{P}-\frac{e}{c}\vec{A})^2 \rangle&=&m^2 \langle
\dot{z}(t)\dot{z}^{\dagger}(t) \rangle -m\hbar\omega_c\nonumber\\
&=&2mkT\nonumber\\
&+&\frac{m\hbar\omega_0 ^2}{\pi}\sum_{j=1}^3 \bigg\{ q_j
\psi(1+\frac{\lambda_j}{\nu})+q^{\prime}_j
\psi(1+\frac{\lambda^{\prime}_j}{\nu})\bigg\}\nonumber\\
&+&\frac{m\hbar}{\pi}\sum_{j=1}^3 \bigg\{p_j
\psi(1+\frac{\lambda_j}{\nu})+p^{\prime}_j
\psi(1+\frac{\lambda^{\prime}_j}{\nu})\bigg\}~.\nonumber\\
\end{eqnarray}
 In the limit of a vanishing magnetic field, the two average values which we
calculate are similiar to the result obtained for a damped harmonic oscillator,
as given by Weiss\cite{1}, of course with a different degree of freedom.

The internal energy can be obtained as
\begin{equation}
 E(\omega_0)=\langle \mathcal{H} \rangle =\frac{1}{2}m \langle
\dot{z}\dot{z}^{\dagger} \rangle -\frac{1}{2}\hbar\omega_c +\frac{1}{2}m\omega_0
^2 \langle zz^{\dagger} \rangle~.
\end{equation}
Taking the derivative with respect to temperature, we find 
\begin{eqnarray}
 C^{Einstein}_{\omega_0 \ne0} &=&2k_B \nonumber\\
& -&k_B \beta \frac{\hbar \omega_0 ^2}{\pi} \sum_{j=1}^3 \bigg\{q_j
\frac{\lambda_j}{\nu}\psi^{\prime}(1+\frac{\lambda_j}{\nu})\nonumber\\
&+&q^{\prime}_j
\frac{\lambda^{\prime}_j}{\nu}\psi^{\prime}(1+\frac{\lambda^{\prime}_j}{\nu}
)\bigg\}\nonumber\\
&-&k_B \beta \frac{\hbar}{2\pi}\sum_{j=1}^3 \bigg\{p_j
\frac{\lambda_j}{\nu}\psi^{\prime}(1+\frac{\lambda_j}{\nu})\nonumber\\
&+&p^{\prime}_j
\frac{\lambda^{\prime}_j}{\nu}\psi^{\prime}(1+\frac{\lambda^{\prime}_j}{\nu}
)\bigg\}~,\nonumber\\
\end{eqnarray}
where $\psi^{\prime}(z)$ are the trigamma functions and $k_B$ is the Boltzmann
constant. 
Finally employing the recurrence formula for trigamma functions leads to 
\begin{eqnarray}
 \psi^{\prime}(1+z)&=&\psi^{\prime}(z)-\frac{1}{z^2}~, ~~~{\rm {and~
also}}\nonumber\\
\sum_{j=1}^3
\{\frac{p_j}{\lambda_j}+\frac{p^{\prime}_j}{\lambda^{\prime}_j}\}&=&0~,
\nonumber\\
\sum_{j=1}^3
\{\frac{q_j}{\lambda_j}+\frac{q^{\prime}_j}{\lambda^{\prime}_j}\}&=&-\frac{1}{
\omega_0 ^2}~,
\end{eqnarray}
from which we obtain Eq.(72).
\section{Einstein Approach ($\omega_0 =0$)}
\vskip 1cm
Here we provide details of the calculations for the case of $\omega_0 =0$. Here,
one of the three roots, viz., $\lambda_1$ vanishes and we are left with just two
roots. From the Vieta equations given in Eq.(47), we can write the new equations
for this particular case as $\lambda_2 +\lambda_3 =\omega_D +i\omega_c ~,
\lambda_2 \lambda_3 =\omega_D (\gamma+i\omega_c)$. 
In the limit of vanishing harmonic oscillator frequency, the energy is obtained
as  Eq.(78)
\begin{equation}
 E=-\frac{1}{2}\hbar\omega_c +\frac{1}{2}m\lim_{t\rightarrow\infty}[\zeta_2
(t)]_{\omega_0 =0}~.
\end{equation}
We can write $ \zeta_2 (t)= \langle
\dot{z}\dot{z}^{\dagger} \rangle$ as
\begin{eqnarray}
 \lim_{t\rightarrow\infty}[\zeta_2
(t)]_{\omega_0 =0} &=&\frac{2k_B
T}{m}+\frac{\hbar\omega_c}{m} \nonumber\\
&+&\frac{\hbar}{m\pi}\bigg\{p_2 \psi(1+\frac{\lambda_2}{\nu})+p_3
\psi(1+\frac{\lambda_3}{\nu})\nonumber\\
&+&p^{\prime}_2 \psi(1+\frac{\lambda^{\prime}_2}{\nu})+p^{\prime}_3
\psi(1+\frac{\lambda^{\prime}_3}{\nu})\bigg\}~.
\end{eqnarray} where
\begin{eqnarray}
 p_2 &=&\frac{[\gamma\omega_D -i\omega_c (\lambda_2 -\omega_D)]}{(\lambda_2
-\lambda_3)}~{\rm{and}}~,\nonumber\\
p_3 &=&-\frac{[\gamma\omega_D -i\omega_c (\lambda_3 -\omega_D)]}{(\lambda_2
-\lambda_3)}~.
\end{eqnarray}
The primed roots are calculated from complex conjugates. 
Hence, the internal energy is
\begin{eqnarray}
 E(\omega_0 =0)&=&k_B T \nonumber\\
&+&\frac{\hbar}{2\pi}\bigg\{p_2 \psi(1+\frac{\lambda_2}{\nu})+p_3
\psi(1+\frac{\lambda_3}{\nu})\nonumber\\
&+&p^{\prime}_2 \psi(1+\frac{\lambda^{\prime}_2}{\nu})+p^{\prime}_3
\psi(1+\frac{\lambda^{\prime}_3}{\nu})\bigg\}~.
\end{eqnarray}
Correspondingly, the specific heat becomes 
\begin{eqnarray}
 C_{\omega_0 =0} ^{Einstein}&=&-k_B \nonumber\\
&-&k_B \beta \frac{\hbar}{2\pi}\bigg\{p_2 \frac{\lambda_2}{\nu}
\psi^{\prime}(\frac{\lambda_2}{\nu})+p_3 \frac{\lambda_3}{\nu}
\psi^{\prime}(\frac{\lambda_3}{\nu})\nonumber\\
&+&p^{\prime}_2 \frac{\lambda^{\prime}_2}{\nu}
\psi^{\prime}(\frac{\lambda^{\prime}_2}{\nu})+p^{\prime}_3
\frac{\lambda^{\prime}_3}{\nu}
\psi^{\prime}(\frac{\lambda^{\prime}_3}{\nu})\bigg\}~.
\end{eqnarray}
This form of the specific heat has been used in the text as the basis of our
discussions of the low and high temperature limits, via Eqs.(80) and (81).

\end{document}